# Plasma Surface Metallurgy of Materials Based on Double Glow Discharge Phenomenon

Zhong Xu[1,], Jun Huang[2, a)], Zaifeng Xu[1], Xiaoping Liu[1], Hongyan Wu[3]

[1] Research Institute of Surface Engineering, Taiyuan University of Technology, Taiyuan 030024, China.
[2] School of Materials Science and Engineering, Nanchang Hangkong University, Nanchang 330063, China
[3] Department of Material Physics，Nanjing University of Information Science and Technology, Nanjing 210044, China

[a)] Author to whom correspondence should be addressed: huangjun@nchu.edu.cn and davidzxu@126.com

## Abstract

Plasma Surface Metallurgy /Alloying is a kind of surface metallurgy/alloying to employ low temperature plasma produced by glow discharge to diffuse alloying elements into the surface of substrate material to form an alloy layer. The first plasma surface metallurgy technology is plasma nitriding invented by German scientist Dr. Bernard Berghuas in 1930. He was the first person to apply glow discharge to realize the surface alloying.

In order to break the limitation of plasma nitriding technology, which can only be applied to a few non-metallic gaseous elements such as nitrogen, carbon, sulfur, the "Double Glow Discharge Phenomenon" was found in 1978. Based on this phenomenon the "Double Glow Plasma Surface Metallurgy Technology", also known as the "Xu-Tec Process" was invented in 1980. It can utilize any chemical elements in the periodic table including solid metallic, gas non-metallic elements and their combination to realize plasma surface alloying, hence greatly expanded the field of surface alloying. Countless surface alloys with high hardness, wear resistance and corrosion resistance, such as high speed steels, nickel base alloys and burn resistant alloys have been produced on the surfaces of a variety of materials. This technology may greatly improve the surface properties of metal materials, comprehensively improve the quality of mechanical products, save a lot of precious alloy elements for human beings.

Based on the plasma nitriding technology, the Xu-Tec Process has opened up a new material engineering field of "Plasma Surface Metallurgy". This Review Article briefly presents the history of glow discharge and surface alloying, double glow discharge phenomenon, basic principle and current status of Double Glow Plasma Surface Metallurgy/Alloying. Industrial applications, advantages and future potential of the Xu-Tec process are also presented.

## 1. Introduction

In 1913, the nitriding process was invented by Adolph Machlet and Adolf Fry of Germany and obtained the US patent.[1,2]

In 1930, Bernhard Berghaus of Germany invented plasma nitriding and obtained German and British patents[3-5]. The invention of plasma nitriding technology is considered as "one of the most important achievements in the field of material surface engineering"[6], and "the cornerstone of modern plasma surface engineering"[7].

Based on the research of plasma nitriding technology for many years, Professor Zhong Xu discovered the "Double Glow Discharge Phenomenon" in 1978, followed by the invention of the "Double Glow Plasma Surface Metallurgy Technology" by using this phenomenon in 1980. The technology can be employed for all of the chemical elements in the periodic table, including solid metal elements and gaseous nonmetallic elements to realize surface alloying on the surface of metallic material. A large number of experimental results have been proved that this method can produce countless surface alloys with high hardness, wear resistance and corrosion resistance on the surfaces of steels, titanium and titanium alloy, copper and copper alloy, intermetallic compounds, and other conductive materials. That turns lower grade materials into high-quality materials. This technology has been patented in the United States, the United Kingdom, Canada, Australia, Sweden and Japan. We have successfully applied this technology in carbon steel plate, sawing tools, colloid mill, chemical valves and other industrial products, and achieved good economic results.

In addition, this technique can also form a gradient alloy ceramic layer (transition from metal substrate to ceramic) on the surface of a metal material. Recently, The Xu-Tec Process, has successfully formed grapheme, Er-$ZrO_2$, ZnO and other luminescent materials on the surface of materials.

The Xu-Tec technology is a typical physical process. Its entire process includes the vacuum, glow discharge, ion bombardment, deposition, diffusion, etc. There is no pollution at all. In addition, the surface alloy layer thickness is only about 0.1-1% of the thickness of the matrix work-piece, hence there is very little alloy elements consumption, which is just about 1% or less of the whole work-piece. It is an environment-friendly and resource-saving technology.

## 2. Material Surface and Surface Technology

In the long-term production practice, people are more and more aware of the importance of material surface. The surface is just as important as the human skin, and it is a barrier to prevent external environmental damage (such as friction, impact of force, corrosion, oxidation, etc.). A large number of facts have proved that the failure and damage of materials often start from the surface. In many working conditions, the requirements for the matrix and surface properties of materials are quite different. Generally speaking, the matrix is required to have high strength and certain toughness, while the surface is required to have high hardness, wear resistance, corrosion resistance, high temperature oxidation resistance, etc. In order to resolve the contradiction between matrix and surface in performance requirements, many surface technologies have been invented to improve the hardness, wear resistance and corrosion resistance of material surface.

The existing surface technology can be basically divided into the following two categories: coating technology and surface alloying technology. Coating technology includes electroplating, deposition, coating, thermal spraying, physical vapor deposition, chemical vapor deposition and so on. It is characterized by mechanical combination between coating and substrate, which is easy to peel and produces cracks under the working conditions of external force and temperature change. The characteristic of surface alloying is that the alloy elements enter the matrix through thermal diffusion and form the surface alloy layer. This creates a very good metallurgical bonding between the surface alloy layer and the matrix. The composition of the surface alloy changes gradiently.

Ion implantation is a kind of special surface alloying technology. It is to force high energy ions (such as bullets) into the surface of the matrix material to form the surface alloy layer. The thickness of the alloy layer is generally less than 1 μm.

Working conditions of mechanical parts (such as gears and shafts) are often affected by external forces, such as stretching, bending, impact, etc., as well as friction, wear, corrosion, temperature change resistance, etc. Therefore, for mechanical parts, surface alloying technology must be used to strengthen their surface properties.

In recent hundred years, in order to resolve the contradiction between the performance requirements of metal parts surface and matrix, and to improve the surface hardness and wear resistance, carburizing and nitriding are the two most important surface alloying technologies. However, how to infiltrate a large number of solid alloy elements into metal materials is a problem that has not been solved.

For a long time, solid alloy elements have been infiltrated into metal materials by solid, liquid and gas methods, such as solid chromizing, liquid aluminizing, etc. However, these methods are very backward with poor working conditions, pollution, low efficiency, and can only be carried out workshop type single piece small batch production. In recent years, the surface alloying technologies of laser beam, ion beam and electron beam have been developed, but they have to be line scanning. The equipments are expensive, energy consumption is high, efficiency is low, therefore, cost is high. All of the above surface alloying technologies have significant shortcomings and limitations. Therefore, how to diffuse a large number of solid alloy elements into metal materials is a problem.

The double glow plasma surface metallurgy technology has greatly filled the gap in the field of surface alloying. It is a very significant breakthrough in the field of surface alloying. This technology will be more and more widely used in metal materials and machinery manufacturing industry. It will save a lot of precious alloy elements and realize huge economic benefits.

## 3. Glow Discharge and Its Characteristics

## 3.1 Glow Discharge

Glow discharge is a very popular phenomenon in our life. When we walk in the night on the city streets, we see very beautiful colorful neon advertising signs everywhere. The light emitted by the neon tube is from the glow discharge.

It was not until the 18th century man began to understood thunder and lightning scientifically. It has been recognized that lightning and lightning are gas discharge phenomena occurring in the atmosphere. Until the glow discharge phenomenon was observed in the vacuum tube of the diode in the laboratory, people realized the association with the flash in the atmosphere, thus uncovering the mysterious veil of the discharge in the atmosphere.

The following are some major milestones connected with glow discharge:
- In 1775, American scientist B. Franklin was first person to study the lightning in the atmosphere. After many experiments, it was confirmed that lightning in the sky is a discharge phenomenon.
- In 1835, M. Faraday, a British scientist, discovered gas glow discharge in a vacuum tube and observed the stratified phenomenon in the glow discharge.
- In 1879, British scientist W. Crookes studied the gas discharge characteristics and called the fourth state of matter as "Plasma".
- In 1928, American scientist I. Langmuir discovered the anode column in discharge is neutral, as densities of the electrons and positive ions are mostly the same. He formally named this state in the anode column as "Plasma".

Since then, glow discharge phenomenon and its application have been explored and studied for more than a century. The plasma produced by glow discharge, as the fourth form of substance (besides solid, liquid and gas) with "positive energy", has gradually been understood. At the same time, it has also been widely used for the development of new technologies and the preparation and development of new materials.

## 3.2 Characteristics of Glow Discharge

As is seen in the Figure 1, no visible light can be observed in the vacuum diode at the very low voltage and current (the dotted line in the curve). As the voltage and current change, it enters Townsend discharge and normal glow discharge, then into abnormal glow discharge. The "abnormal glow discharge" region is used for Plasma nitriding process. Within the abnormal region, any increase in voltage, gives an increase in the current density, while the current density is uniform around the entire cathode surface which is indicated by a uniform glow. As the voltage is further raised it would approach towards the glow discharge/arc discharge transition threshold point, where the glow discharge collapses and high current density arcs form. It is essential to control cathode discharge for avoiding arcing.

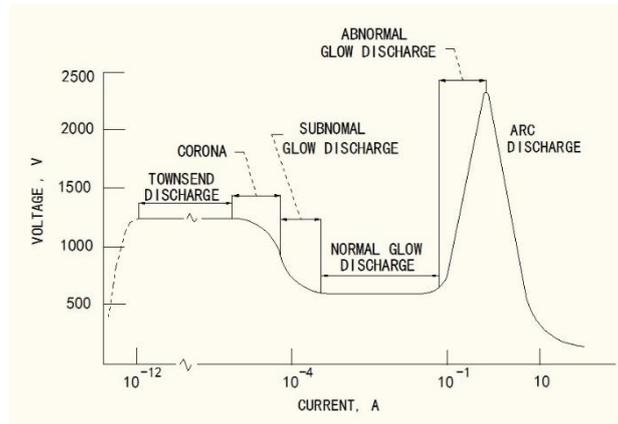

FIG. 1. Voltage-Current characteristics of different types of discharge in Argon.

The arc discharge is partial discharge, confined in a small area and would cause the work-piece ablation, further induce equipment damage. Therefore, in the plasma nitriding process, we must try to avoid the occurrence of arc discharge.

### 3.3 Interaction between Ions and Material Surface

The basic physical and chemical interactions between ions and surface of substrate material are schematically shown in Figure 2.

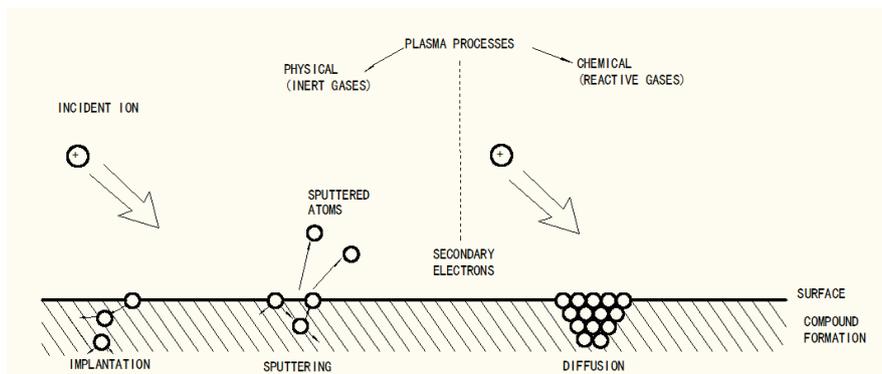

FIG. 2. Ion-Surface interactions.

The positive ions generated in the glow discharge, such as non-metallic element argon and nitrogen, as well as metal elements, driven by the electric field, will bombard the cathode surface, resulting in some interaction effects as are listed below:

1. Sputtering: Any material with negative potential in the vacuum chamber will be sputtered by positive incident ions. We can use any negative electrode made of alloying element as a target for the alloying element supply.
2. Heating: The incident ions with high energy bombard the surface of substrate and releases energy to the substrate. The negative cathode can be heated by ion bombardment. Temperature of the work-piece can be higher than 1200 degree.

3. Implantation: Any energetic ion can be implanted into the substrate surface to form an implantation layer.
4. Diffusion: If the work-piece is heated by ion bombardment to enough temperature for diffusion, the incident ions are also diffuse deeply into the surface of the work-piece to form a much greater thickness alloying layer than that of implantation layer.
5. Cleaning: The ions bombarding the cathode surface can clean away the oil stain and oxides decomposition on the work object surface.
6. Etching: Ion bombardment can erode and strip off the surface layer of the cathode material. This effect has been used as an important method for the glow discharge analyzer of surface composition distribution.
7. Creating: Ion bombardment can create a defect layer including a large number of vacancies and dislocations of the surface of the cathode, which can be conducive for dissolve more alloying elements and diffusion of alloying element.
8. Accelerating chemical reaction: Glow discharge greatly increases the speed of chemical reaction and reduces the chemical reaction temperature.

## 3.4 Stratified Phenomenon of Glow Discharge

In glow discharge, the stratification phenomenon of bright space and dark space is a very important phenomenon. Figure 3 shows the typical Stratified Phenomenon of Glow Discharge. From the cathode to the anode, there exist several glow discharge spaces: Aston dark space, Cathode glow, Cathode dark, Negative glow, Faraday space, Positive column, Anode glow, and Anode dark space. The first four spaces, including Aston dark, cathode glow cathode dark, and negative glow are very important regions for maintaining glow discharge. The total discharge width Dk of these four zones is also called the width of cathode potential drop region.

When the anode moves toward cathode, only the width of the positive column is reduced, but other parts remain unchanged.   However, once the anode cathode comes in the negative glow region, all of the glow discharge will be immediately extinguished. The higher the gas pressure is, the smaller the Dk value would be. This is a very important phenomenon for all of the technologies related to glow discharge.

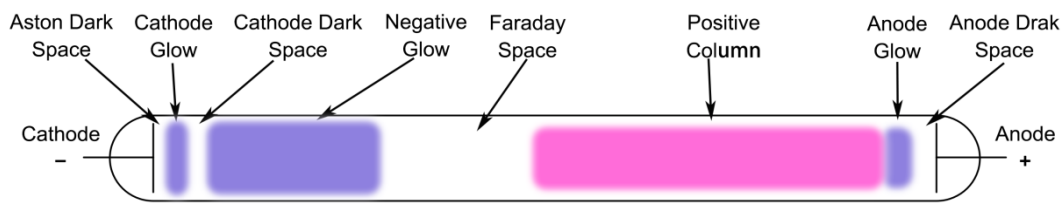

FIG. 3. Stratified Phenomenon of Glow Discharge.

The cathode potential drop zone Dk is a very important parameter in glow discharge. It plays an important role in the design and manufacture of equipment related to the application of glow discharge phenomena; especially all the gap protection of the power transmission device of the cathode must be based on Dk value. In addition, Dk value should also be taken into account in the design of part shape and the placement of parts in glow discharge equipment; otherwise, glow discharge can easily be converted to arc discharge, resulting in the burning and destroy of parts and glow equipment.

## 4. Double Glow Discharge Phenomenon

### 4.1 Discovery of Double Glow Discharge Phenomenon

Since 1972, we began to study ion nitriding technology. In order to further increase the speed of ion nitriding, we developed "titanium ion nitriding". In order to improve the surface hardness of steel materials, we also developed "Ti-N-C ternary ion nitriding technology". This led us to whether we can use glow discharge to diffuse the solid metal elements into the surface of steel materials to form a surface alloy layer.

As we observe the experimental phenomena of ion nitriding, we explored the way to infiltrate the solid alloy elements into the material surface. We realized that, in order to accomplish the surface alloying under glow discharge condition by using solid metal elements, the key is to vaporize the solid metal elements, so that the glow discharge gas contains the atoms of the solid metal elements.

In the experimental observation, we noticed sparks and/or local micro arc discharges emitted from the surface of the working piece. In addition, we also observed that there are more and more steel fine powders on the stove chassis. It had made us to realize that the solid metal elements in cathode electrode would be sputtered by positive ion bombardment from plasma under the glow discharge condition. The sputtered off metallic atomic species move into the glow discharge space, then deposit on the surface of the stove chassis.

This observation led us to the conclusion that we can use the ion bombardment and sputtering to realize gasification of solid alloy elements. We managed to set up a second cathode (as a source electrode, made of the desired alloying elements) between the anode and cathode in plasma nitriding equipment. Driven by two DC power supplies, two glow discharge (plasma) zones would be established respectively between the anode and cathode as well as the anode and second cathode. We called this state as "Double Glow Discharge Phenomenon". Ion bombardment at the second cathode makes the desired solid alloying element to be sputtered and gasified into glow discharge space.

The Double Glow Discharge Phenomenon is shown in Figure 4. In a vacuum chamber there are three electrodes: the grounded anode, the cathode and an additional negative electrode. Two DC power supplies are applied to the cathode and the second negatively charged electrode separately. The power supplies provide an output voltage from 0 to 1200 V by using a silicon-controlled rectifier. The vacuum chamber is first pumped to a base pressure below 0.1 Pa and back-filled with pure argon gas to a process pressure of 10 Pa to 100 Pa. Under the electric field induced by two high voltage power supplies, argon gas will be electrically broken down and ionized, so that two sets of plasma zones are generated. Both the cathode and the second negatively charged electrode are surrounded by the glow discharges, Therefore, it is called as "Double Glow Discharge Phenomenon".

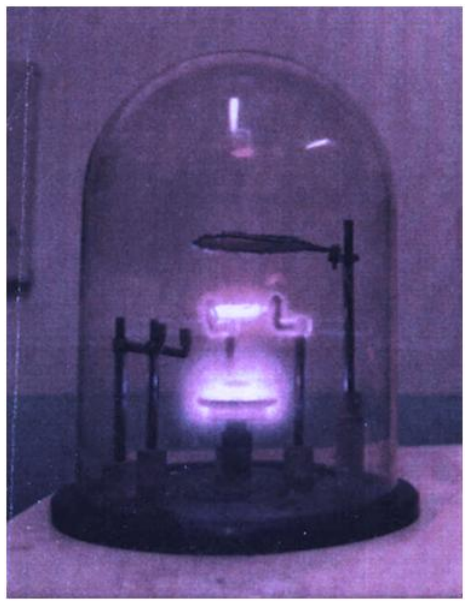

FIG. 4. Double glow discharge phenomenon.

In 1980, the "Double Glow Discharge Phenomenon" led us to the invention of the "Double Glow Plasma Surface Alloying Technology". We discovered that the Double Glow Plasma Surface Alloying Technology can be applied to any solid chemical element such as tungsten, molybdenum, nickel, chromium, etc. and their combination to conduct surface alloying. Surface alloys with gradient concentration of alloying elements have been produced on the surfaces of steels, titanium alloys and inter-metallic compounds and so on.

**4.2 Double Glow Hollow Cathode Discharge (DG-HCD)**

Hollow Cathode Discharge was first discovered by Paschen, a German experimental physicist, in 1916. Hollow cathode discharge is a special discharge mode in glow discharge with very high current. It is widely used in light source, vacuum coating, vacuum metallurgy, spectral analysis and so on.

The Double Glow Hollow Cathode Discharge (DG-HCD) is formed by two sets of cathodes (i.e., the work-piece and the source) driven with two different electric potentials. The experimental device is shown in Figure 5 and set in a sealed vacuum container, including the first cathode (Z2) and the 2$^{nd}$ cathode (Z1) (as is known as the source electrode), two power supplies (V1 and V2). Two cathodes made of a low-carbon steel plate are placed in parallel with a relative distance adjustable in the range of 10-l00mm. The supply output voltage can range of 0-1000V. Working discharge gas is industrial-grade pure argon and working gas pressure ranges in 10-130Pa.

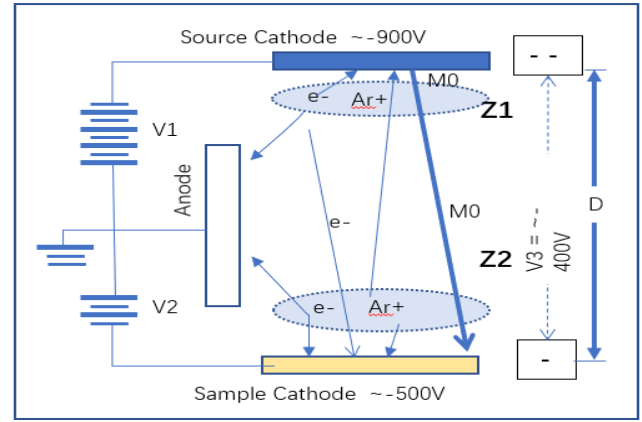

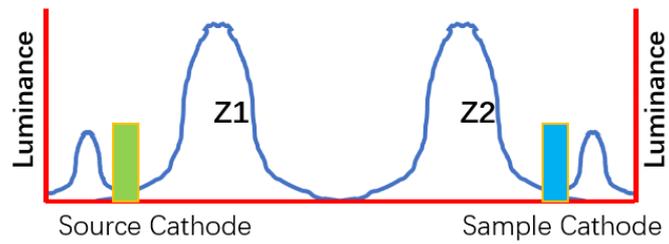

(a)

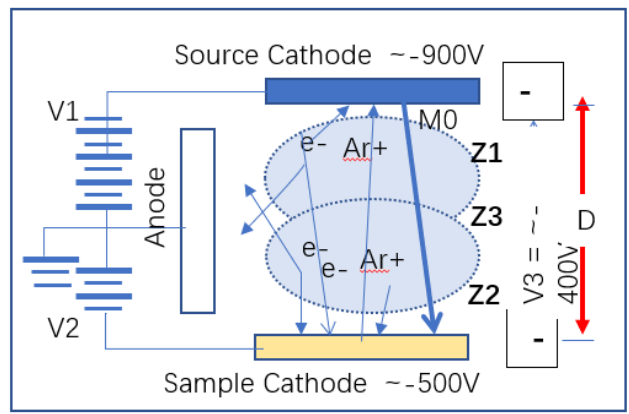

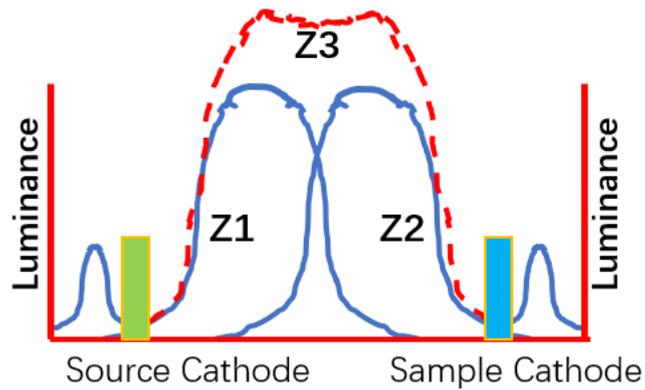

(b)

FIG. 5. Experiment setup for double glow discharge, (a) intensity of double glow discharge. Z1 and Z2 on Source and Samples cathodes and no intersection between source and sample. (b) Z3 is the total glow discharge intensity when two cathodes get closed and the glow discharge zones overlap.

When two power supplies (V1 and V2) are switched on, the glow discharges are generated along the surface of cathode and sample respectively. At first, the negative glow regions of the two cathodes are well defined and do not intersect as shown by the curve Z1and Z2 in Figure 5(a). Double glow discharge can also produce hollow cathode discharge as shown as Figure 5(b). As argon pressure decreases, the thickness of the negative glow region increases. When the negative glow regions of the source Z1 and the sample Z2 are overlapped with each other in the space between two cathodes, the glow brightness is significantly enhanced. If the pressure is reduced further or the two cathode voltages are increased, the two cathode glow regions are mutual overlap and cross, the brightness of the glow discharge and two cathodes current density will greatly increase shown as curve Z3. This is the Double Glow Hollow Cathode Discharge (DG-HCD). Since the discharge potentials of the two cathodes are not equal, we call this phenomenon also as the unequal potential hollow cathode discharge.

The current amplification effect of the double glow hollow cathode discharge is shown in Figure 6 [5]. When the cathode voltage $U_c$ is increasing to 400V, then both the sample cathode current $I_c$ and source electrode current $I_s$ will sharply increase.

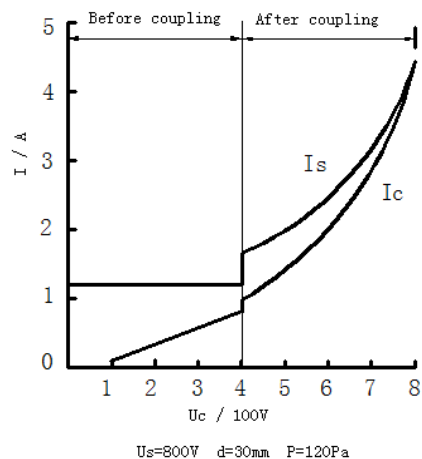

FIG. 6. Current amplification effect of unequal DG-HCD

The luminous intensity and current density of double glow hollow cathode discharge can be 1 to 3 orders of magnitude higher than that of common glow discharge.

## 5. Double Glow Plasma Surface Metallurgy Technology

The "Double Glow Plasma Surface Metallurgy Technology", also known as the Xu-Tec process, is a major breakthrough to break the restriction of plasma nitriding and covered by US and international patents.[8]

### 5.1 Basic Principle
The Xu-Tec Process uses double glow discharge to introduce solid alloying elements to the surface of conductive or semi-conductive materials for subsequent diffusion into the substrate to form a surface alloy. As sketched in Figure 7, there are three electrodes in a vacuum chamber: ground anode, negatively biased cathode (substrate), and negatively biased source electrode, which works as a sputtering target made of the desired alloying element. The source electrode can be fabricated with rod, plate, and cylinder according to the geometry of the work-piece. For industrial furnace, usually alloy plates made by powder metallurgy are adopted for the convenience of the composition adjustment and durability. Two DC power supplies are employed to control the discharge on the two negative electrodes separately. The power supplies provide an output voltage from 0 to 1200 V by silicon-controlled rectifier.

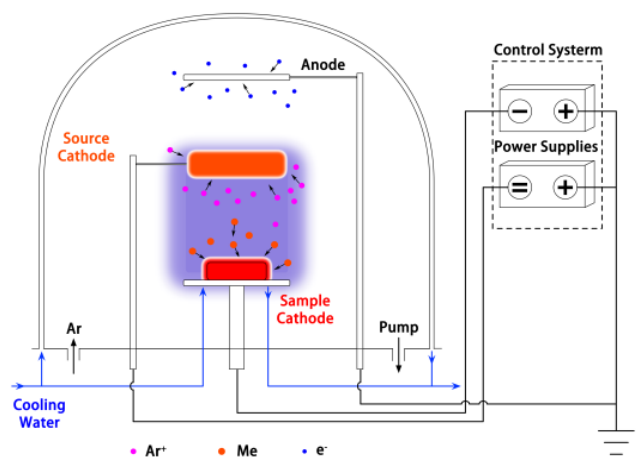

FIG. 7. Principle of Xu-Tec process.

During the Xu-Tec process, the vacuum chamber is first pumped to 0.1 Pa and back-filled with pure argon. With the two power supplies turned on, argon gas will break down into positive ions and electrons and both the source electrode and the substrate are surrounded by the glows.

In most cases, double glow plasma surface metallurgy technology is carried out in the mode of double glow hollow cathode discharge. The high strength of hollow cathode discharge and high current density will greatly increase the sputtering strength of ion bombardment and the supply of alloy elements.

The working gas medium used in the double glow plasma surface metallugy technology is argon. Under glow discharge conditions, argon atoms will be excited and ionized during collisions with particles. Excitation is that the inner electrons of the argon atom are excited to jump to the outer layer with a high energy potential, and become unstable excited electrons. When such a high-energy electron jumps back to a low-energy inner layer due to instability, the energy released by the electron appears in the form of light, which is the main source of glow. Ionization refers to the collision of argon atoms with other particles to cause their electrons to escape from the argon atoms, forming a positively charged argon ion and a free electron. Driven by an electric field, positive ions of argon bombard the work-piece and source electrode with negative potential. The work-piece is heated to a high temperature due to the bombardment, and the source element is sputtered out due to the ion bombardment. In the space of the glow discharge, the atomic elements of the alloy that are sputtered out, like the argon atoms, will also undergo the activation and ionization. In a DC electric field, the positive ions that are constantly accelerating pass through the collision, and can also make other neutral particles become fast-moving particles. Under the bombardment of ions, some atomic groups composed of a small number of atoms may also be sputtered in the workpiece and source material. It can be seen that in the glow discharge space formed by the anode, cathode (workpiece) and source, there are positive ions of argon, positive ions and electrons of various alloy elements and matrix elements, neutral particles of various elements and Fast particles and various charged or uncharged atomic groups. Therefore, there are many types and shapes of particles in the glow discharge space, and the interaction between them is more complicated.

As you can imagine, the physical image in the glow discharge space is very complicated. However, under the action of the electric field, all negatively charged particles are directed to the anode, all positively charged particles are directed to the workpiece and the source, and neutral particles are free to move randomly. However, the positively charged ions and particles that move at high energy and speed will also make multiple collisions and drive other particles to move quickly during their directional movement. The experimental measurement has proved that the ionization rate (ratio of the number of positive ions to the number of all particles) under the glow discharge condition is very low, and most of them are neutral particles and particle clusters. However, the movement of neutral particles and atomic clusters will be affected by positive ions moving at high speed in directional motion, which will change the speed and direction.

The positive ions of argon bombard the source, on the one hand, heating the source, and on the other hand, causing alloy elements in the source material to be sputtered out, passing through the glow discharge space, moving toward the workpiece and adsorbing on the surface of the workpiece. Because the workpiece is heated to a high temperature that can diffuse metal elements under the action of ion bombardment, the alloy elements adsorbed on the surface of the workpiece diffuse into the interior of the workpiece surface to form a surface alloy layer.

The Xu-Tec technology is a typical physical process. Its entire process includes the vacuum, glow discharge, ion bombardment, deposition, diffusion, etc. There is no pollution at all. It is an environment-friendly and resource-saving technology.

The Surface Metallurgy/Alloying technology is different from Plating，Coating, Deposition Spray, PVD, CVD and Thin film technology. The composition of the surface alloy layer formed by diffusion process of surface metallurgy has gradient distribution between surface alloy and matrix material. Therefore, the bond between the surface alloy and the matrix is very strong[9.].

October 2017, the German Springer Press published a monograph with the title "PLASMA SURFACE METALLURGY with Double Glow Discharge Technology - Xu-Tec Process" detailing the "Double Glow Plasma Surface Metallurgy/Alloying Technology".[9]

## 5.2 Diffusion Mechanism Under Ion Bombardment

Compared with that of the conventional solid, liquid and gas surface alloying technologies, the formation speed of the alloy layer here is generally 1-3 times higher, and the depth of diffusion or the thickness of the alloying layer is much large.

To explain the higher diffusivity of plasma nitriding, professor Xu proposed a new diffusion mechanism about high concentration vacancy defect layer which is formed on the substrate surface by ion bombardment in 1979. In the double glow plasma surface alloying process, it was further proposed that "Vacancy Gradient" at the surface forms a diffusion channel for surface absorbed alloying elements to diffuse into the substrate.

In 1988, they further expanded the vacancy diffusion mechanism and some diffusion model under ion bombardment. The vacancy diffusion mechanism under the condition of ion bombardment can be summarized as follows:

1. Ion bombardment can effectively remove oil stain and oxides on the surface of the work-piece. After the working-piece surface cleaning, the exposed fresh and activated surface is beneficial to adsorb incoming alloying elements. Alloying elements diffuse into the work-piece without any barriers.
2. Under the bombardment of large number positive ions, a defect layer will be formed on the surface of the work-piece with a large number of vacancies and dislocations. The defect layer will dissolve more alloying element, so that the solubility of alloying elements at the substrate surface would exceed the solubility of the alloying element in the alloy phase diagram. Therefore, a higher concentration gradient formed on the surface will accelerate the diffusion process of alloying elements into the working-piece at high temperature.
3. Much high vacancy concentration in the surface defect layer will enhance the vacancy diffusion into the matrix and a vacancy gradient layer will be formed. The high vacancy concentration gradient at the surface would provide a fast path for the alloying elements to diffuse into the matrix.

The above points suggested that the Double Glow Plasma Surface Metallurgy is faster than other traditional surface alloying processes. However, these views are only hypotheses and inferences, which need to be proved by experiments.

## 5.3 Processing Parameters

There are many technological parameters for double glow plasma surface metallurgy technology, such as argon pressure, cathode voltage, source electrode voltage, currents of cathode and source, distance between substrate and source, temperature, holding time and so on.

According to a large number of optimizing experiments, we can summarize the typical process parameter ranges are as follows:
- Argon gas pressure: 25-50 Pa
- Source voltage: 800-1200 V
- Work-piece voltage: 300-600 V
- The distance between the work-piece and the source: 10-30 mm

- Alloying temperature: 900-1000℃
- Holding time: 1-10 hours

In the actual operation of the surface alloying, we usually set argon gas pressure at 35Pa, the distance between the work-piece and the source at 20mm, and the alloying temperature ~~is~~ at 1000 degrees. The holding time depends on the thickness of the alloy layer. The work-piece voltage and source voltage are the main adjustable operating parameters.

In most cases, the double glow hollow cathode discharge mode is used for double glow plasma surface alloying. In this kind of a discharge mode, all kinds of discharge parameters influence each other, and the interaction is complicated.

## 6. General summarize of Research Results

According to incomplete statistics, there are 954 published articles on double glow plasma surface metallurgy technology, including 662 SCI articles. This tutorial can only very simply introduce some representative research results as examples and show the functions and features of the Xu-Tec process.

In the early experimental studies, Xu-Tec process was mainly applied to ferrous metals. A series of surface alloys[10-13] have been formed on the surfaces of plain carbon steel by plasma surface alloying by using single element of W, Mo, Cr, Al, Ti, Zr, Pt, Ta and multi-elements of W-Mo, Ni-Cr, W-Mo-Ti, Ni-Cr-Mo-Nb, etc..[14-20] A variety of functional alloyed layers were developed to improve the wear resistance, corrosion resistance and oxidation resistance of the substrate. For example, high-speed steel and nickel-based alloys are formed on the surface of steel materials.[21,22] The depth of alloy layer can vary from several microns to 1 mm, and the concentration range of alloy elements is from 1% to 100%. In the last decade, there are many new types of alloyed layers, such as Nb, Mo-B, Fe-Al-Nb, Fe-Al-Cr-Nb, and Cu-Ni, used to improve the wear resistance of the steels.[23-27] In addition, many alloyed layers are also applied on the surface of the steels to improve corrosion resistance and oxidation resistance, such as the Ni-based alloyed layer,[28-33] intermetallic alloyed layer,[34-36] ,Al-Cr alloyed layer,[37] and alumina coatings.[38-41]

Since 1997, the research and application of plasma surface metallurgy have been extended to Ti and Ti alloys, copper and copper alloys, TiAl, $Ti_2AlNb$ intermetallic compounds, etc.

In terms of titanium and titanium alloys, the surface alloys of Mo, Zr, borided, silicide, Ti-Mo, Mo-Ni, Mo-N, Mo-C, Mo-Cr, and Ti-Mo-N have been formed to improve the tribological properties of the substrates.[42-51] Qiu Zhongkai[52] deposited Zr and Zr-Er alloyed layers onto the TC11 alloy. The Zr and Zr-Er alloyed layers showed not only high mechanical property, but also excellent corrosion resistance. With the application of titanium alloys in aero engine, we began to study how to improve the oxidation resistance of titanium alloys. Wei Dongbo[53-57] prepared Cr and CrNi alloys on Ti6Al4V alloy surface to improve its oxidation resistance. Moreover, Xu-Tec process was also used to improve the corrosion resistance of Ti-6Al-4V employed in fuel cells. Xu Jiang[58-62] fabricated a serial of nanocomposite coatings, such as ZrC, ZrCN, TaN, and TiSiN coatings, on metallic bipolar plates, employed in proton exchange membrane fuel cells, to meet the needs of corrosion resistance and electrically conductivity. These coatings, which exhibited excellent corrosion resistance, electro-conductivity and hydrophobicity, could protect metallic bipolar plates from corrosive attack. It is also reported that Nb, $TiO_2$, and Ta alloyed layers were prepared on pure titanium surface by double glow plasma surface alloying technology.[63-67]

In the aspect of copper and copper alloys, wear-resistant alloys such as W, Ti, Cu-Ti, Cu-Ni, and TiN/Ti were studied experimentally.[68-71] Wear-resistant alloys of Nb, Mo and Cr have been studied experimentally in TiAl and Ti2AlNb metal compounds. It was reported that many different alloyed layers, such as Mo, Cr, Cr-W and Ni-Cr alloyed layers, formed on $Ti_2AlNb$-based alloys to improve their wear resistance and high temperature oxidation resistance.[72-76] Chen Xiaohu[77] synthesized a novel Cr/CrC multilayer on γ-TiAl alloy. The multilayer reduced the friction coefficients and improved the oxidation resistance of γ-TiAl. At the same time, there were also many scholars enhanced wear resistance and oxidation resistance of the γ-TiAl alloys by preparing W-Mo, MoSiAlY, and CrCoNiAlTiY alloyed layers.[78-80] For niobium alloys, the surface alloys of Ir, Mo, Fe-Cr-Mo-Si have been studied experimentally. Chen Zhaofeng[81-87] prepared Ir-Zr and Ir coatings on the surface of Ti, Mo, and Nb substrates to improve their oxidation resistances at high temperature. The growth mechanism of the coatings was also studied. In addition, surface alloys with special physical and chemical properties are formed on the surfaces of molybdenum, tungsten, tantalum, Fe-Al-Cr and C/C materials. The preparation of surface alloys on composite materials by Xu-Tec process have also been reported. Chen Zhaofeng[88-92] prepared W, Ir and W/Ir multilayers onto C/C composites and graphite to protect the composites from quick oxidizing when ablated by an oxyacetylene torch at 2000 ℃. Meanwhile, they also prepared multilayer Ir coating on WC to protect substrates at 2000-2200 ℃ in oxidizing flame.[93] In order to avoid hydrogen embrittlement in carburizing process, a new double glow non-hydrogen carburizing process has been successfully studied.

In 2003, we put forward the research direction of ceramization of metal materials and metallization of ceramic surface. The former is to improve the bonding strength between ceramics and matrix materials, while the latter is to solve the welding problems of metals and ceramics. TiN, WC, TiC, Ti (CN) gradient ceramics have been successfully formed on the surface of ceramics, which greatly improves the bonding strength between ceramics and metal matrix.[94-97] Metal elements Ti, Zr, W and Ta layers are formed on the surface of diamond, 3Y-TZP, TiSi30, and Si3N4 ceramics, which solves the welding problem between diamond or ceramics and other materials.[98-100]

By combining diffusion and deposition processes, a biphasic surface consisting of a coating and a diffusion layer can be achieved. The transition zone is between metal and ceramic. An example of mixed modified layer is TiN/Ti formed on the surface of the low carbon steel.[101,102] Recently, Professor Wu Hongyan successfully formed graphene, Er-$ZrO_2$, ZnO and other luminescent materials on the surface of materials using Xu-Tec Process.[103-107]

In summary, these above experimental results are just as examples to illustrate the wide application of the Xu-Tec process. In fact, this technology can form countless surface alloys on the surfaces of a wide range of matrix materials. The Xu-Tec process transforms low-grade materials into high-quality materials. Therefore, there should be countless industrial applications with significant economic impact.

The technology has been successfully applied to carbon steel plate, sawing tools, colloid mill, chemical valves and industrial products for special purposes. It solves many technical problems and achieves good economic benefits. In recent years, Xu-Tec process has been used to form alloys with special properties, such as the antimicrobial properties of stainless steel surface, and many other research results.[108]

In conclusion, a large number of experimental results have shown that Xu-Tec technology is a very powerful surface alloying technology, and has a very broad application prospects in metal materials and mechanical manufacturing industry.

## 7. Plasma Surface Metallurgy of carbon steels

Some typical microstructures and compositions of samples alloyed by Xu-Tec process with a single element are illustrated in Figure 8-10, respectively.

Figure 8a shows an optical cross-sectional micrograph of the 1020 steel after tungsten alloying treatment at 1000 ℃ for 5 hrs. It can be seen that a W-alloyed layer of about 60 μm thick is formed compactly and closely bonding to the steel substrate. The EDS composition analysis of tungsten as a function of the depth from the surface is shown in Figure 6b. The tungsten concentration is about 19 wt % near the surface; then it drops to about 1.6 wt % near 60 μm. This confirmed the diffusion is indeed about 60 μm. Few carbide precipitates in the W-alloyed layer does not change the hardness after alloying process. Through changing processing parameters, a two-layer structure of W-coating/W-diffusion may be made on carbon steels. The tungsten coating can be used for tribological applications, especially for those at elevated temperatures due to its high melting point and good wear resistance.

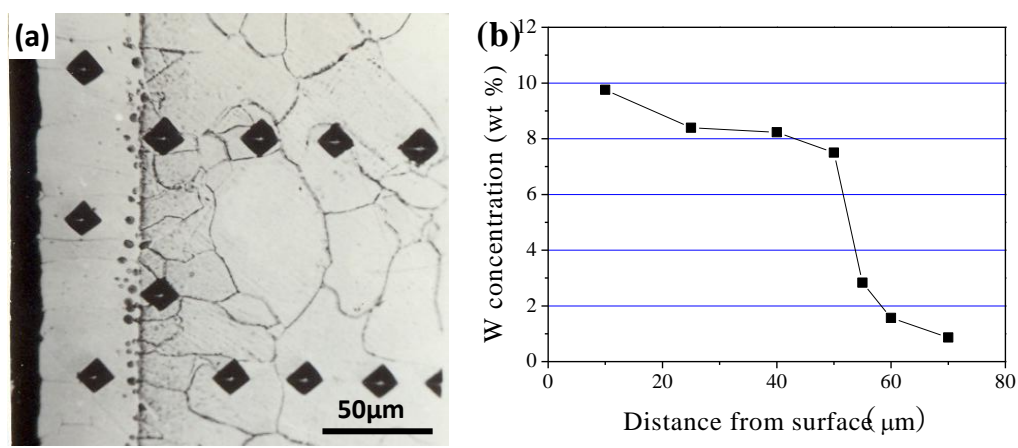

FIG. 8. (a) Cross-sectional micrograph of steel 1020 after plasma tungsten alloying at 1000 ℃ for 5 hrs and (b) tungsten concentration profiling by EDS analysis.

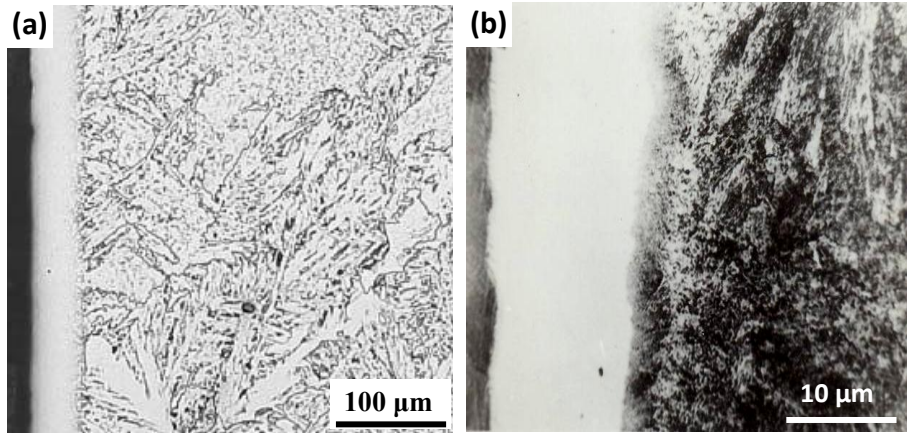

FIG. 9. Micrograph of (a) 1020 and (b) 1080 steels after Cr-alloying at 1000 ℃ for 2 hrs.

Figure 9 is the optical micrograph of the 1020 and 1080 steels treated by chromium alloying at 1000 ℃ for 2 hrs. The Cr-alloyed layer obtained on the low carbon steel is obviously thicker than that on the high carbon steel.

Figure 10a and 10b show the micrograph and Al-distribution of 1045 steel after plasma alloying of aluminum. on the 1045 steel at 950 ℃ for 3 hrs.

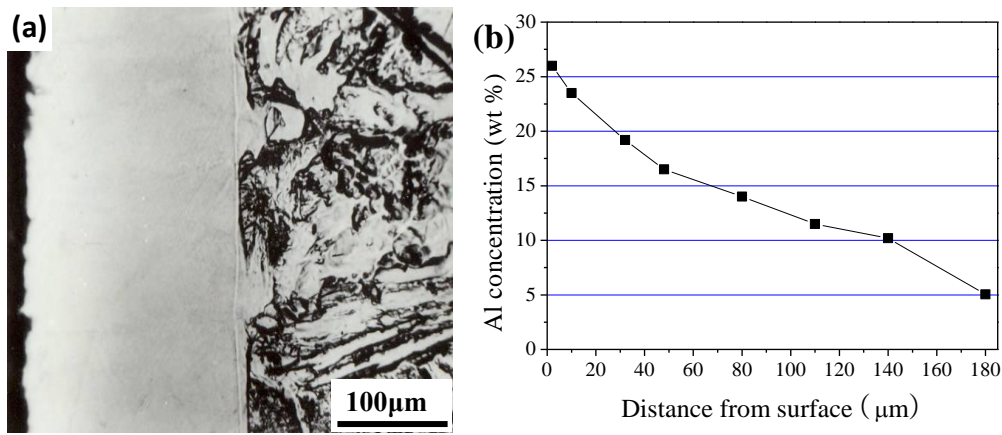

FIG. 10. (a)Micrograph of 1045 steel after plasma aluminum alloying and (b) aluminum concentration profiling by EDS analysis.

## 8. Plasma Surface Metallurgy of titanium and titanium alloys

Since many important alloying elements, such as tungsten, molybdenum, vanadium, niobium, zirconium, tantalum, hafnium, etc., have large solubility with titanium, it makes the Xu-Tec Process ideal in surface alloying of titanium alloy. Molybdenizing on titanium and its alloys is attractive, mainly by its excellent wear resistance of Mo-modified alloy layers.

As is shown in Figure 11, a thick Ti-Pd alloy formed on titanium surface after plasma palladium alloying at 900 ℃ characterizes a multi-layer structure of Pd-deposit/TiPd2+Ti2Pd/α-Ti(Pd) layers.

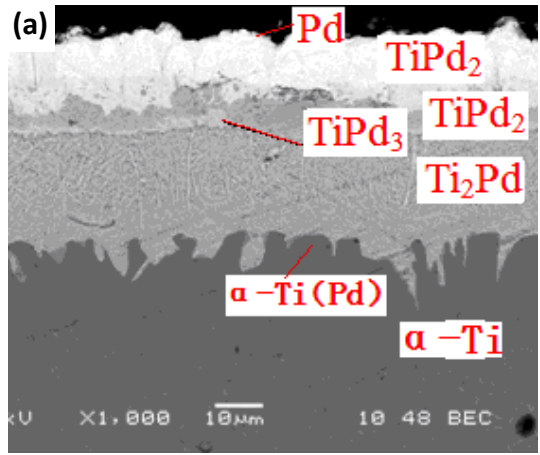

FIG. 11. Micrograph of Pd-alloyed layer formed on titanium.

## 9. Plasma Surface Metallurgy of intermetallic alloys

Figure 12a shows the SEM morphology of Nb-alloyed layer formed on TiAl at 1100 ℃ for 3 hrs. Figure 12b shows its composition distribution obtained by GDOES, the alloyed layer is divided into niobium-rich/aluminum-rich/titanium-rich/substrate.

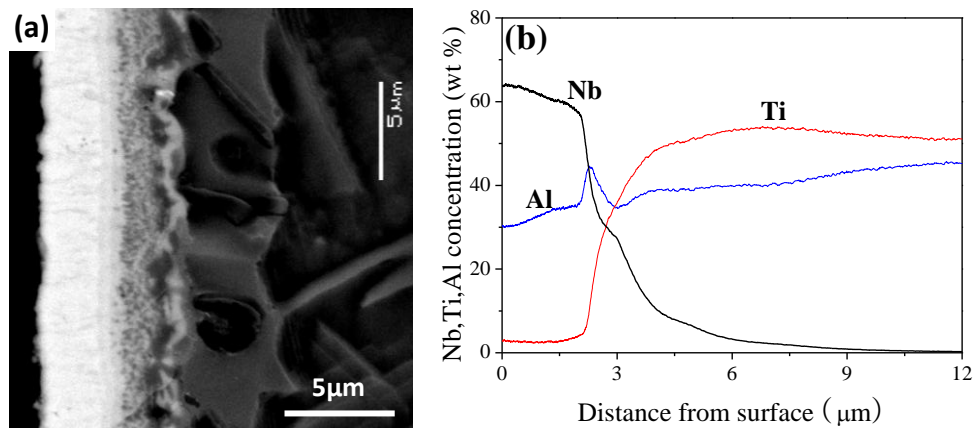

FIG. 12. (a) Cross-sectional micrograph of TiAl alloy after plasma niobium alloying and (b) concentration profiling by GDOES analysis.

## 10. Gradient ceramization of metal surface

After Ti-alloying, Ti-N co-alloying is carried out by Xu-Tec process, the composite layer of TiN/Ti is formed. The microstructure and Ti-N distribution of plasma surface TiN/Ti alloyed layer are shown in Figure 13. The thickness of the Ti alloyed layer is about 14 μm and the content of Ti and N decreases gradient in depth as a diffusion behavior.

It is very obvious that the surface alloyed layer is a typical gradient ceramic layer which ensures that metallurgical bonding with good bond strength is produced in the interface between the alloy layer and the substrate.

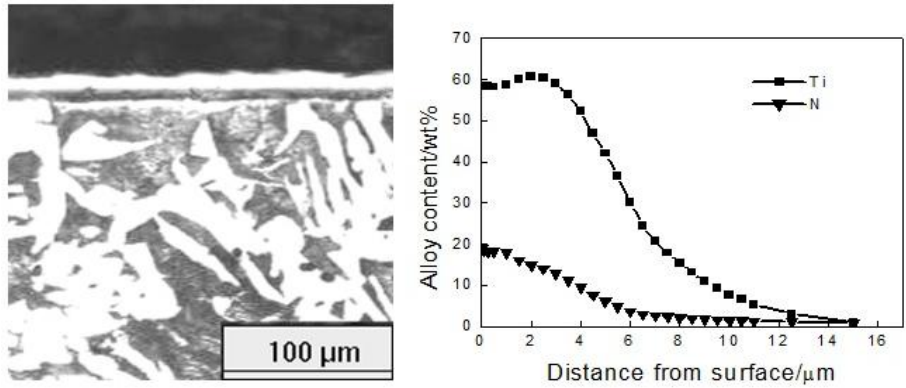

FIG. 13. (a) Microstructure and (b) Ti, N distribution of a Ti-N co-diffusion layer on steel 1045.

## 11. Metallization of ceramic surface

Besides metal materials, the ceramic surface can also be metallized by Xu-Tec process. A micrograph of the brazed joint as shown in Figure 14, the ceramic $Si_3N_4$ is surface titanizing by Xu-Tec process, then braze-welded through Ag-Cu alloy and bonds with carbon steel. The strength of the joint can reach 205 MPa. In addition, using zirconium source materials, the TiSi3O substrate is alloyed to form a composite of Zr-coating/Zr-Si diffusion layer, as is shown in Figure 12. This double alloyed layer is then oxidized into stable $ZrO_2$/$ZrO_2$+$SiO_2$ mixture, forming a duplex ceramics with better corrosion resistant on the TiSi3O surface.

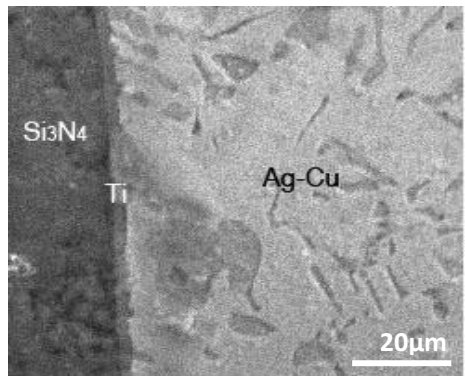

FIG. 14. Cross section micrograph of Si3N4 brazed joint.

## 12. Plasma Surface Metallurgy Nickel Base Alloy

### 12.1 Plasma Surface Ni-Cr Alloy

Ni-Cr alloy layer was formed on carbon steel surface by using double glow plasma Ni-Cr alloying with Ni80-Cr20 as source materials. The thickness of Ni-Cr alloyed layer was 70μm. The content ratio of Ni-Cr follows that of Ni80-Cr20 target. The total content of Ni and Cr of the surface alloy reaches 95wt% at the surface, with a diffusion gradient decreasing deeply into the matrix. Its corrosion resistance has reached or preceded austenite stainless steel.

## 12.2 Plasma surface Ni-Cr-Mo-Nb Super-alloys

Using Inconel 625 as the source material, plasma surface alloying processes were conducted on surfaces of industrial pure iron, steel 1010, and stainless steel 304 at 850~1000℃ for 3 hrs.. The results show that a surface alloyed layer with element composition similar to that of Inconel-625 can be formed on these three substrates.

Figure 15 shows the microstructure and alloying element (Ni, Cr, Mo, Nb) depth distribution of surface alloyed layer on industrial pure iron. Nickel based alloyed layer is uniform, continuous and dense. A strong metallurgical bonding is formed between the surface alloyed layer and the substrate. The thickness of the alloyed layer is more than 40μm. Ni, Cr, Mo, Nb content gradient decreases in sub-surface layers.

The corrosion resistance of plasma surface metallurgy Ni-Cr-Mo-Nb alloy has been investigated by electrochemical method in 3.5%NaCl, 5%HCl and 5%$H_2SO_4$ solution. The surface alloy on pure iron has similar corrosion resistance to Inconel 625. The corrosion resistance of surface alloy on steel 1010 is lower than that of Inconel 625, but much better than that of stainless steel 304.[109]

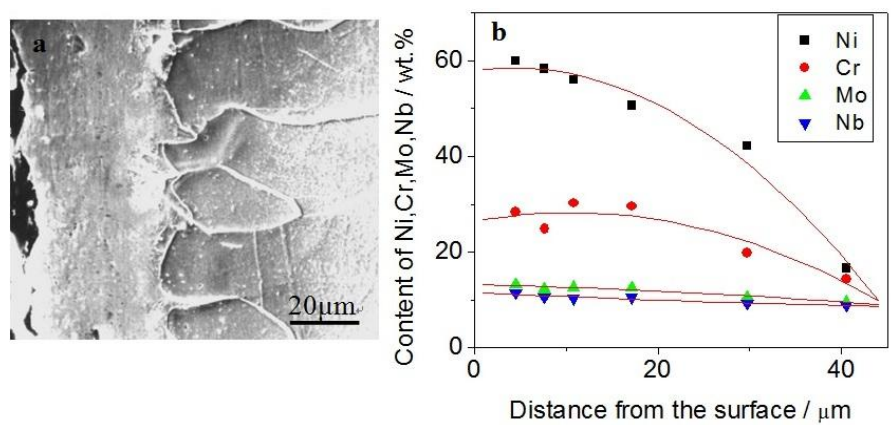

FIG. 15. (a) Microstructure and (b) Ni, Cr, Mo, Nb distribution of alloyed layer on iron.

## 12.3 Plasma surface Ni-Cr-Mo-Cu Super-alloys

Using the Hastelloy C-2000 alloy as the source material, Ni-Cr-Mo-Cu alloying was conducted on the surfaces of steel 1020 and stainless steel 304 at 900~1000℃.

Figure 16 shows the microstructures and Ni, Cr, Mo, Cu distribution curves of alloyed layer on steel-1020. Surface composition of an alloyed layer is similar to Hastelloy C-2000, the total content of alloy elements surpasses 90wt.%. The average thickness of alloyed layer is more than 35μm after 3hrs alloying. Content of alloy element in alloyed layer of steel 1020 drops from the surface to the substrate matrix. A good metallurgical bonding is formed between the alloyed layer and the substrate.

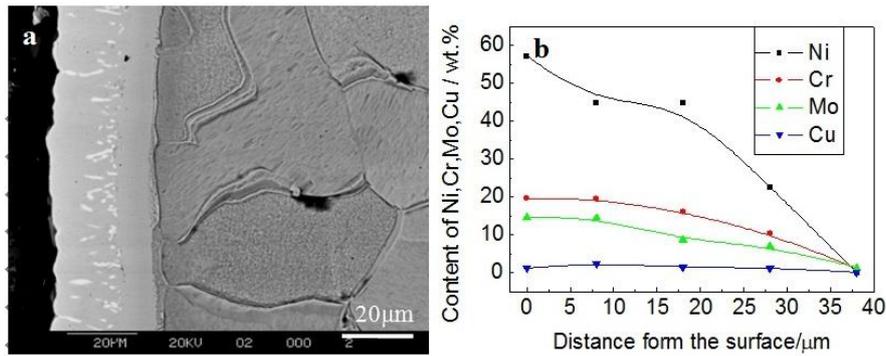

FIG. 16. (a) Microstructure and (b) Ni, Cr, Mo, Cu distribution of plasma alloyed layer on steel 1020.

## 13. Plasma Surface Metallurgy High Speed Steel

Plasma surface metallurgy high speed steel (HSS) is another typical application of Xu-Tec process.

Through plasma W-Mo alloying and carburizing a Fe-W-Mo surface alloy layer which is similar to HSS can be formed on surface of plain carbon steel. Figure 17 shows the microstructure and alloying element content of alloyed layer (W, Mo) on steel 1020 after double glow plasma W-Mo alloying at 1050℃ for 8 hrs. The thickness of W-Mo alloyed layer is over 300μm. The element content on the surface is about 3wt.% for W and 10wt.% for Mo, respectively. W, Mo content gradients decrease along the depth of the alloyed layer. After carburizing, amount of fine carbides dispersedly distribute in the alloyed layer.

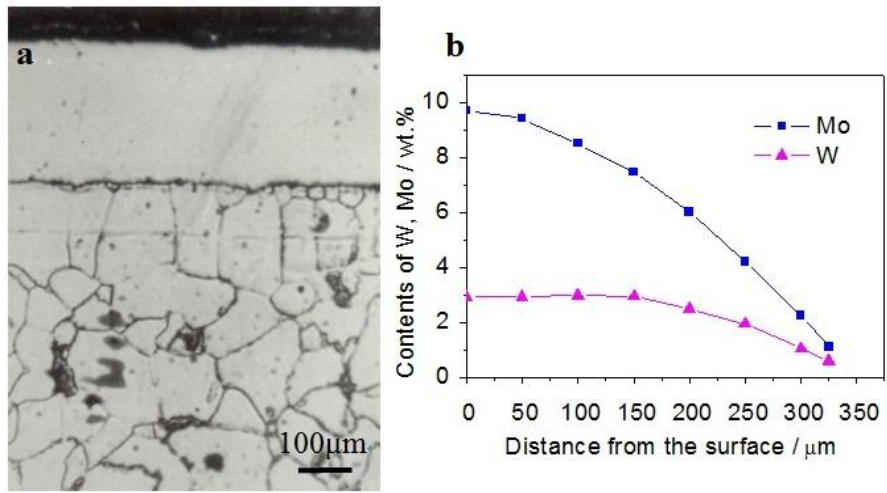

FIG. 17. (a) Microstructure and (b) W, Mo distribution of plasma alloyed layer on steel 1020.

## 14. Plasma Surface Metallurgy Ni-Cr Alloy Steel Plate

The first product by application of the Xu-Tec process was the plasma surface metallurgy Ni-Cr alloy plate. Our goal was to develop this new kind of plate to partly replace the conventional stainless steel plate. [110,111]

For this reason, we specially designed and manufactured a large double glow industrial furnace for the carbon steel plate. The internal structure of Xu-Tec industrial furnace for steel plate is shown in Figure 18. In the center of the furnace body, two flat Ni80Cr20 alloy plates are set up in parallel as source electrodes. The work-piece of a carbon steel plate is placed between the two source electrodes. This setup is able to provide uniform temperature and sputtering yield. Its simplicity in manufacture and arrangement helps to achieve long service time. After plasma surface alloying, the Ni and Cr contents on the surface of the carbon steel plate maintain at an 80% and 20% ratio.

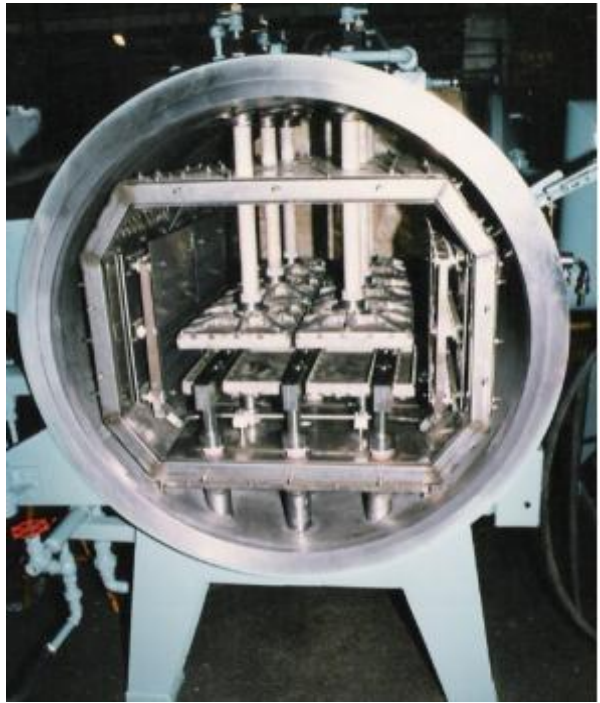

FIG. 18. Internal Structure of Xu-Tec Industrial Furnace for Steel Plate.

We successfully prepared nickel-based alloy layer on the surface of low carbon steel plate with a size of 1 m long and 0.5 m wide. Figure 19 shows the cold rolled steel plate before (left side) and after (right side) the double glow plasma surface metallurgy treatment.

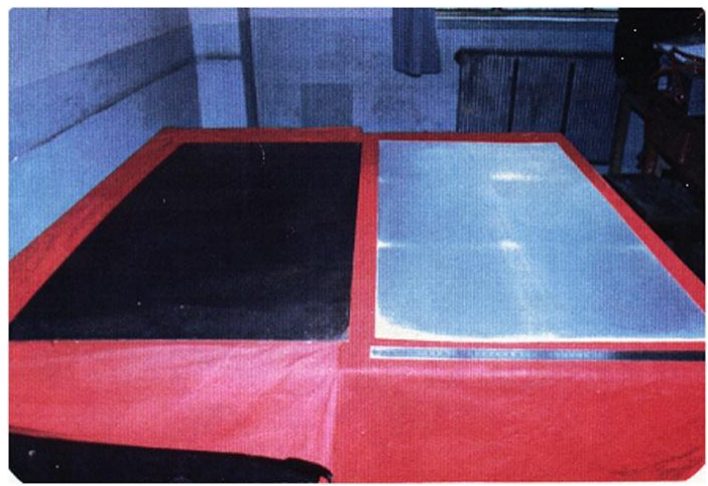

FIG. 19. Cold rolled low carbon steel plate before(left) and after (right) Xu-Tec process.

The Plasma Surface Metallurgy Stainness Steel Plate may be the most important application of the Xu-Tec process.

"The formation of stainless steel on the surface of ordinary carbon steel is one dream of the world's metallurgical industry for long Times."

The Xu-Tec process had been used to form Nickel base alloys with much better corrosion resistance than stainless steel on the surface of low carbon steel plate. China produced a total of 25.77 million tons of stainless steel in 2017, of which 85% was in a form of plates. If the plasma surface metallurgy stainless steel 314 plates replace 20% of stainless steel plate, about 0.85 million tons of Nickel and 1.03 million tons of Chromium could be saved.

If 20-50 % of the world's stainless steel plates are replaced by this new type of plasma surface metallurgical stainless steel plates, the huge amount of precious alloy elements such as nickel and chromium will be saved and the economic value would be enormous.

## 15. Plasma Surface Metallurgy Chemical Valves and Flanges

Many parts of chemical and petroleum industries devices, such as valve body, fixing plate, valve cover, flange, etc., require use of stainless steel materials. It is a good opportunity for us to use ordinary carbon steels with plasma surface metallurgy with Ni-Cr alloying treatment. Its service life and corrosion resistance would be even better than pure stainless steel, and the cost would be only 1/3 of stainless steel. Figure 20 shows the flange parts and steel plate after plasma surface metallurgy Ni-Cr alloying treatment.

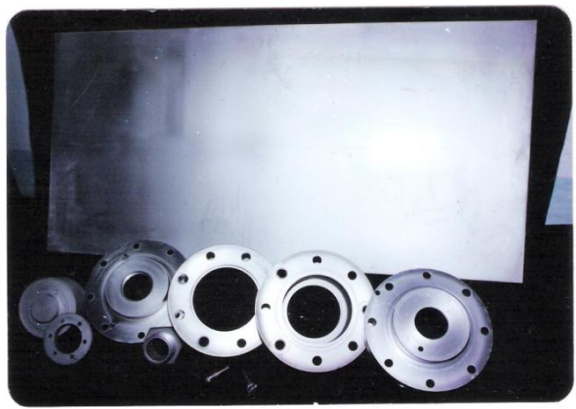

FIG. 20. Flange and steel plate after plasma surface alloying with Ni-Cr.

## 16. Plasma Surface Metallurgy High Speed Steel Saw Blade

The plasma surface metallurgy high speed steel saw blade is a new product by the application of the Xu-Tec process. It is made of low carbon steels or low alloy steel and alloyed by the Xu-Tec process with molybdenum and tungsten, and followed by carbonizing to form an alloyed layer which is similar to high speed steel. Then, the blade is treated by subsequent quenching and tempering. These novel Xu-Tec HSS saw blades have an excellent cutting performance, good plasticity and toughness. They can also be flexibly bent. The Xu-Tec saw blade can perform as well as bimetal HSS saw blade.

According the British BS 1919-1983 standard, nine cold-rolled 18-8 stainless steel strips, by 25 mm wide and 2.6 mm thick are used to compose cutting test bar with 25x23.4 mm cross section. The total cumulative time of cutting test bar of the hand saw blade for ten times shall not exceed 124 minutes, and the wear rate must be less than 40. Numerous cutting test results of the Xu-Tec HSS saw blade indicate that the saw blades meet the British Standard Specification for Bimetal Saw Blade.[112,113]

Xu-Tec HSS handsaw blades are produced in a horizontal furnace, as shown in Figure 21. A frame is used for hanging the handsaw blades. There are ten rows of handsaw blades are placed in the furnace and each row consists of 1000 pieces of the handsaw blades, five rows of Mo80W20 alloy plate of 6 mm thickness are hung in the upper part of the furnace as source materials. The teeth of handsaw blade face the W-Mo alloy plate. The space between the teeth and the source plate is about 20mm. 10000 pieces of handsaw blades can be alloyed at one time in each furnace.

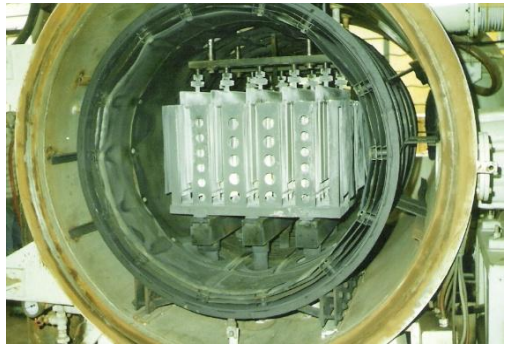

FIG. 21. Internal structure of Xu-Tec industrial furnace treating saw blade.

Figure 22 shows the microstructure of such a W-Mo alloying layer after plasma surface alloying with W- Mo and carburizing, where an alloyed HSS layer was formed. After quenching and tempering, the surface microstructure of the blade is HSS tempered martensite. The contents of W-Mo on the surface of the prong are 3% and 10%, respectively. The diffusion depth at diagonal line of the teeth has reached above 300 μm.

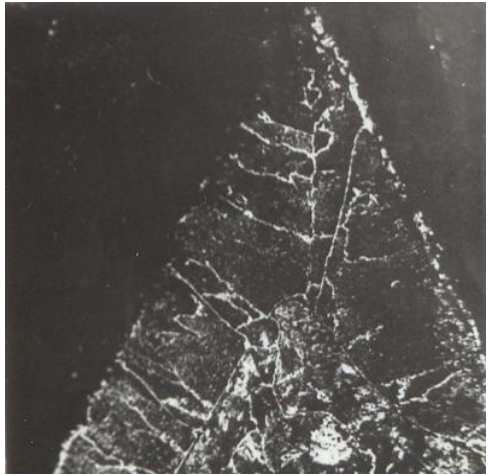

FIG. 22. Microstructure of prong of Xu-Tec HSS saw blade.

The carbide grain structure of M2 HSS and Xu-Tec HSS are shown in Figure 23 (a) and (b), respectively. The carbides of Xu-Tec HSS are very fine and disperse. Absence of coarse primary carbides in Xu-Tec gives the biggest advantage over M2-HSS formed during the liquid solidification and after much hot rolling.

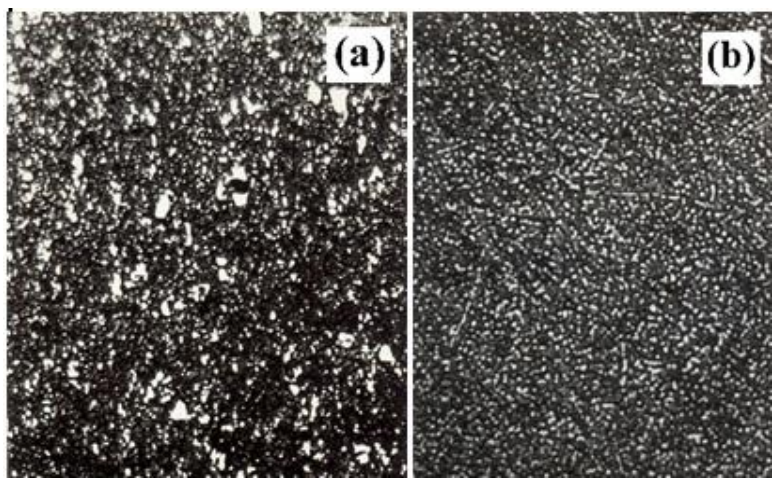

FIG. 23. Morphology and distribution of carbides in: a M2 HSS; b Xu-Tec HSS.

It should be emphasized that some Xu-Tec HSS handsaw blades demonstrated excellent cutting performance during the cutting test. They were able to cut off the stainless steel test bar 100 times. The time consumed in the last 10 times reached 124 minutes and the wear rate was less than 40. This result is probably due to the fine (smaller than 2 microns) and dispersed carbides in Xu-Tec HSS handsaw blade. The Xu-Tec HSS handsaw blade may have a great potential for future optimization development.

If we say that the traditional high-speed steel is the first generation of high-speed steel, the powder metallurgy high-speed steel is the second generation of high-speed steel, and the plasma surface metallurgy high-speed steel is the third generation of high speed steel.

## 17. Plasma Surface Metallurgy colloid mill

Another application of the Xu-Tec process is the surface modification of colloid mill. Mill grinders, made of stainless steel 9Cr18, are used in production of ethanol and experience severe wear. Figure 24 shows the colloid mill after plasma surface alloying with W and Mo, carburizing, quenching and tempering. The surface composition of the colloid mill is over 30%wt W, Mo and 2.5%C. The hardness after the heat treatment reaches 63-65HRC.

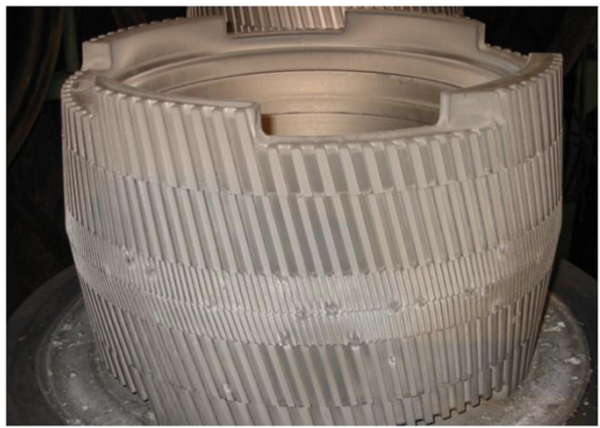

FIG. 24. Colloid mill after Xu-Tec plasma surface alloying, carburiziong and quinching and temprining.

The wear resistance of the W-Mo alloyed colloid mill is greatly improved by double glow plasma surface treatment. Evaluation by the production practice of Langfang General Machinery Co., Ltd. show the average service life of colloid mill has extended for more than five times, from 6-10 days to 50 days.

## 18. Production Line

A pilot production line of plasma surface metallurgy Ni-Cr alloy plate and W-Mo HSS handsaw blade with 800,000 annual output was established at the Taiyuan University of Technology. The production line was composed of the Xu-Tec furnace for plasma surface alloying, HZQT-150 furnace for vacuum carburizing and high pressure gas quenching, HZRD -50 furnace for vacuum tempering and nitriding. The production line is showed in Figure 25.

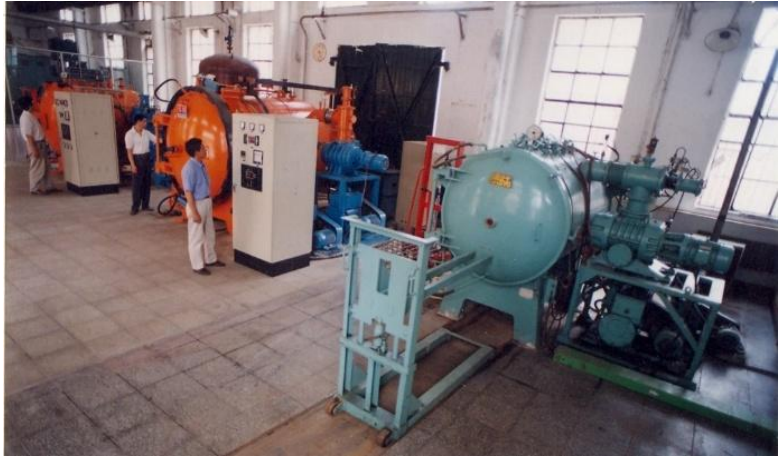
FIG. 25. Pilot production line of plasma surface metallurgy plate and blade.

In recent years, the Xu-Tec Process has been successfully applied in diamond tools, antibacterial stainless steel, plastic mold steel, ceramic piston rings and functional films et.al.

## 19. Advantages of the Xu-Tec Process

Compared with other surface engineering technologies，such as 3-beam (electron, ion, laser) and 3-state (gas, solid, liquid) surface alloying technology，the advantages of the Xu Tec Process are listed below:

1. Universality and Flexibility

Since powder metallurgy alloys can be used to make the source material target, the composition of surface alloys is almost unrestricted. The concentration of the alloying element in the surface alloy can be varied in a wide range from a few percent up to 95 %. The depth of the surface alloy can be formed from a few micron-meters to a few hundred micron-meters.

2. Conserve Natural Resource

The surface alloy layer thickness is only about 0.1-1% of the thickness of the working-piece. The alloy elements consumption is small, just about 0.1-1% of the whole working-piece.

3. Create no Environmental pollution

The Xu-Tec technology is a typical physical process. The entire process of the technology includes the vacuum, glow discharge, ion bombardment, deposition, diffusion, etc., and there is no chemical pollution. It is an environment-friendly technology.

4. Save Energy

Diffusion rate of metal element of this technology is about 1-3 times faster than that of traditional solid, liquid and gas surface alloying. The equipment has no traditional heat treatment furnace insulation materials and components. Heating of the working piece is directly applied by plasma to the local surface. So the plasma surface metallurgy technology can save energy more than 30%.

5. Excellent Bonding Properties

Composition of the alloyed layer gradually decreases with the depth of the work-piece. There exist no distinct boundary between the surface alloy and matrix, hence the risk of layer delamination is greatly reduced or eliminated.

6. Easy Scaling up for Large Area

According to the characteristic of glow discharge, the area of glow can be enlarged with the increase of the area of cathode. Theoretically, the area of the glow discharge is basically unlimited. Today, we have reached the surface alloying area of 6 square meters.

7. Composition of the alloyed layer can be controlled

Through the technological and electrical parameters signals applied to the power supply and computer, the composition can be controlled.

8. Forward looking technology

The technology developed by vacuum, glow discharge, plasma, sputtering, deposition, diffusion and other physical phenomena has distinct foresight, advanced nature and long-term timeliness.

## 20.  Summery and Prospect

1. Plasma Surface Metallurgy belongs to the category of material surface technology and metallurgy technology in material science and engineering. It is a brand-new discipline in the field of Surface Engineering.

2. Double glow plasma surface metallurgy technology (Xu-Tec Process) breaks the limitation that plasma nitriding can only be applied to few non-metallic gas elements. Any element in the periodic table of chemical elements and its combination can be used for the Xu-Tec process to realize surface alloying of conductive materials, which greatly enlarges the scope of surface alloying technology.

This is a forward-looking major basic process technology in the field of metal materials and machinery manufacturing.

3. Numerous experiment results have indicated that almost countless types of surface alloys have been successfully produced by the Xu-Tec Process on the surfaces of carbon steels, titanium alloys and international compounds. The plasma surface metallurgy high speed steel is a third generation of high speed steels, besides chemical metallurgy and powder metallurgy of high speed steel.

A gradient ceramic layer can be also produced on the surfaces of metals and a metal layer can also be prepared on the surfaces of ceramics by the Xu-Tec Process..

4. The Xu-Tec Process has been successfully applied to carbon steel plate, sawing tools, colloid mill, chemical valves and industrial products for special purposes. This technology has solved many technical problems and achieved good economic benefits. It is a very powerful and practical technology and is on the threshold of large scale industrial application.

In a word，this technology transforms low-grade materials into high-quality materials. Hence there should be countless industrial applications with significant economic impact.

5. The Xu-Tec process is a typical physical process. Its entire process includes the vacuum, glow discharge, ion bombardment, sputtering, deposition, diffusion, etc. There is no pollution. In addition, it consumes very little alloy elements since the surface alloy layer thickness is only about 0.1-1% of the thickness of the matrix work-piece. This technology is a truly resource-saving and environment-friendly, and can create huge economic benefits.

6. The plasma Surface Metallurgy/Alloying technology is different from Plating，Coating, Deposition Spray, PVD, CVD and Thin film technology. The composition of the surface alloy layer formed by diffusion process of surface metallurgy has gradient distribution between surface alloy and matrix material. Therefore, the bond between the surface alloy and the matrix is very strong.

7. The Double Glow Plasma Surface Metallurgy is an interdisciplinary technique involving vacuum and gas discharge, sputtering and diffusion, phase transformation, surface physics, mechanical design and computer control, etc. The mechanism and detailed physical image of this technique is not fully understood due to the complexity of the plasma environment and particle interactions. A series of fundamental issues are still unresolved, such as diffusion mechanism under ion bombardment, reactions between alloying elements and chemical elements in matrix, characteristics and stability of the low-temperature glow and arc discharge, etc. In particular, the precise control of surface alloy composition is a subject that needs further study. The solutions to these challenging problems would further release the potential power of this technology.

8. Plasma, as the fourth form of matter, has been widely used in the development of new technologies and the preparation of new materials. However, the application of plasma in metal materials and machinery manufacturing industry is almost blank except plasma nitriding. The Xu-Tec process technology will be an important trend and direction for the future development of metal surface engineering technology.

9. Material science is a science that studies the relationship between the chemical composition, the internal microstructure, the properties of materials and processing technology (production means). It is a science full of Hegel's Philosophy, such as quantitative change and qualitative change, unity of opposites and negation of negation. As a material scientist, we are deeply moved by the dialectics in the material science system.

10. During the long times of research on plasma nitriding and double glow plasma surface metallurgy, we have deeply realized that the role of philosophical thought in guiding scientific research. We deeply realize that Philosophy is an important source of human innovative thinking and activities. We also deeply feel that the role of philosophy is everywhere. Whether you are conscious or unconscious, philosophy is affecting your thinking and activities invisibly.

For example, the dialectical relationship between quantitative change and qualitative change, which is prevalent in nature, is also prevalent in the thinking process of scientific research. When you persevere in thinking about a problem, with the accumulation of time and quantity, an accidental phenomenon or event may enable you to get the key to solving the problem in an instant. This may be what people call "inspiration". For example, in my years of exploring the application of glow discharge phenomena in solid alloy elements to achieve surface alloying, I attach great importance to the observation of glow discharge phenomena in plasma nitriding. Although I observed the glow discharge phenomena in the experiment every day, one day, suddenly because of the spark on the surface of the work-piece and the metal powder in the furnace chassis, I came up with the idea of vaporizing solid alloy elements by ion bombardment sputtering. Thus, I designed and found the phenomenon of double glow discharge. Thereafter, on the basis of experimental study on the characteristics of double glow discharge, the phenomenon was applied to surface alloying of solid alloy elements, thus the "double glow plasma surface metallurgy technology" was invented.

We do indeed to believe that if we use philosophy to guide the study of natural science, we can understand and grasp natural science more profoundly from the whole point of view.

## Acknowledgment

The authors wish to thank their colleagues and all doctorate students for their contributions in the Xu-Tec process.